# Utilizing Astrometric Orbits to Obtain Coronagraphic Images

*John M. Davidson*

*Mail Stop 321-520*
*Jet Propulsion Laboratory,*
*California Institute of Technology*
*Pasadena, CA*

*[John.m.davidson@jpl.nasa.gov](John.m.davidson@jpl.nasa.gov)*

*December 23, 2010*

**Abstract**

We present an approach for utilizing astrometric orbit information to improve the yield of planetary images and spectra from a follow-on direct detection mission. This approach is based on the notion—strictly hypothetical—that if a particular star could be observed continuously, the instrument would in time observe all portions of the habitable zone so that no planet residing therein could be missed. This strategy could not be implemented in any realistic mission scenario. But if an exoplanet's orbit is known from astrometric observation, then it may be possible to plan and schedule a sequence of imaging observations *that is the equivalent of continuous observation.* A series of images—optimally spaced in time—could be recorded to examine contiguous segments of the orbit. In time, all segments would be examined, leading to the inevitable detection of the planet. In this paper, we show how astrometric orbit information can be used to construct such a sequence. We apply this methodology to seven stars taken from the target lists of proposed astrometric and direct detection missions. In addition, we construct this sequence for the Sun-Earth system as it would appear from a distance of 10 pc. In constructing these sequences, we have assumed that the imaging instrument has an inner working angle (IWA) of 75 mas and that the planets are visible whenever they are separated from their host stars by ≥ IWA and are in quarter phase or greater. In addition, we have assumed that the planets orbit at a distance of 1 AU scaled to luminosity and that the inclination of the orbit plane is 60 degrees. For the individual stars in this target pool, we find that the number of observations in this sequence ranges from 2 to 7, representing the *maximum* number of observations required to find the planet. The *probable* number of observations ranges from 1.5 to 3.1. These results suggest that a direct detection mission using astrometric orbits would find all 8 exoplanets in this target pool with a probability of unity and that the *maximum* number of visits required (i.e., the worst case) is 36 visits. The *probable* number of visits is considerably smaller, about 20. This is a dramatic improvement in efficiency over previous methods proposed for utilizing astrometric orbits. We examine how the implementation of this approach is complicated and limited by operational constraints. We find that it can be fully implemented for internal coronagraph and visual nuller missions, with a success rate approaching 100%. External occulter missions will also benefit, but to a lesser degree.

## Introduction and Summary

We present an approach for utilizing astrometric orbit information to improve the yield of planetary images and spectra from a follow-on direct detection mission. This approach is based on the notion—strictly hypothetical—that if a particular star could be observed continuously, the instrument would in time observe all portions of the habitable zone so that no planet residing therein could be missed. This strategy could not be implemented in any realistic mission scenario. But if an exoplanet's orbit is known from astrometric observation, then it may be possible to plan and schedule a sequence of imaging observations that is *the equivalent of continuous observation.* Two conditions are required. First, the planet's orbit parameters must be known, including the period. Second, the instrument's inner working angle (IWA) must be smaller than the star-planet separation at maximum extension. It is not required that the planet's orbital phase be known. Given the first condition, the planet is constrained to lie on an ellipse of known size, eccentricity, and orientation and to be moving about it at a known speed. Given the second, the planet will make two excursions[1] per revolution outside the IWA and the excursions will be of known duration, $\tau$. During each excursion, the planet will be visible to an imaging instrument for time $t_1$, where $t_1 \leq \tau$. From this, it is possible to construct a sequence of optimally spaced observations to examine contiguous segments of the orbit for the presence of the planet. This is a systematic approach that will in time observe all portions of the orbit, leading to the inevitable detection of the planet.

Previous studies (Savransky et al. 2009; Brown 2009) have addressed the utility of astrometry as a precursor to direct detection. Those investigations examined two particularly relevant questions. *(i) Given astrometric orbits, could a direct detection mission plan and schedule its observations to coincide with the periods, possibly of short duration, when the exoplanet is visible? (ii) Given a target list on which every star is known to host a planet, but for which the periods of visibility are not known, could a direct detection mission improve its yield of images and spectra?* In both cases, astrometry was shown to be of limited value. The first approach fell short because it required an extrapolation of orbital phase from an epoch in the distant past in the presence of standard errors. In the second case, the astrometry mission was used only as a classifier or "target finder." No use was made of the orbit information.

In this paper, we address the utility of astrometry from a third perspective. *(iii) Given astrometric orbits, could a direct detection mission plan and schedule a sequence of optimally spaced observations to examine contiguous segments of the orbit, leading to the inevitable detection of the planet?* We construct this sequence for a target pool of seven stars taken from the target lists of proposed astrometric (Catanzarite and Shao 2011) and direct detection (Brown and Soummer 2010) missions. In addition, we construct this sequence for the Sun-Earth system as it would appear from a distance of 10 pc. In constructing these sequences, we assume that the imaging instrument has an inner working angle (IWA) of 75 mas and that the planets are visible whenever they are separated from their host stars by ≥ IWA and are in quarter phase or greater. We assume that the planets orbit at a distance of 1 AU scaled to luminosity and that the inclination of the orbit planes is 60 degrees. For the individual stars in this target pool, we find that the number of observations in this sequence ranges from 2 to 7, representing the *maximum* number of observations required to find the planet. The *probable* number of observations ranges from 1.5 to 3.1. These results suggest that a direct detection mission using astrometric orbits would find all 8 exoplanets in this target pool with a probability of unity and that the *maximum* number of visits required (i.e., the worst case) is 36 visits. The *probable* number of visits is considerably smaller, about 20.

---

[1] *These excursions will be of different length for the general case of a non-circular orbit. For the purpose of this discussion, we assume that the orbit is circular.*

This is a dramatic improvement in efficiency over the previous methods proposed for utilizing astrometric orbits. We attribute this to the facts that we do not require unrealistic orbit accuracy as in case (i) above, nor do we discard orbit information altogether as in case (ii). However, this approach does require that observations of target stars be carried out with a particular cadence. We examine how this last requirement affects the feasibility of our approach. Imaging missions, as with all missions, must function within operational limits. We consider three different coronagraph types—external occulters, internal coronagraphs, and visual nullers. These have differing solar exclusion zones, slew rates, and number of stellar visits allowed in a five-year mission. We find that our approach significantly improves the yield of images for internal coronagraph and visual nuller missions, with a success rate approaching 100%. External occulter missions also benefit, but to a lesser degree

## Utilizing Astrometric Orbits to Obtain Coronagraphic Images

Consider a terrestrial[2] planet orbiting in the habitable zone of its host star (Fig. 1). An astrometry mission has determined its period and three-dimensional orbital parameters. An imaging instrument is pointed at the system. The planet is visible whenever its angular separation from the star is greater than the instrument IWA and is in quarter phase or greater.

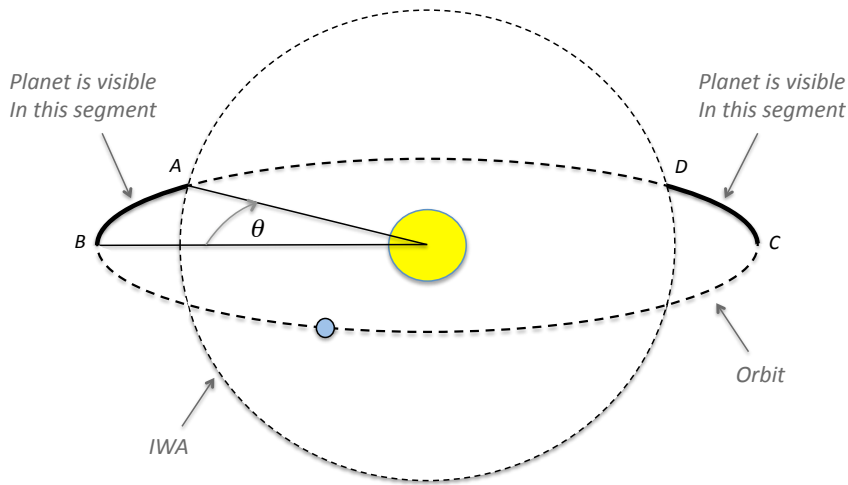

Figure 1. A terrestrial planet orbits in the habitable zone of its host star and is observed by an imaging instrument. The dashed circle represents the projection of the instrument IWA. The planet is visible when it passes through orbit segments AB and CD. The duration of the planet's visibility during any one of these passages is $t_1$.

---

[2] A "terrestrial" planet is taken here to have a mass in the range 0.5—10 $M_\oplus$

The probability, $P_1$, of observing this planet in a randomly scheduled first imaging visit is given by the fraction of the time that it is visible during one orbit. Thus,

$$P_1 = 2\frac{t_1}{T} = \frac{\theta}{\pi} \tag{1}$$

where $t_1$ is the duration of the planet's visibility near maximum extension, $T$ is the period of revolution, and $\theta$ is the difference in phase between point $A$ where the orbit emerges from behind the IWA and point $B$ where the orbit reaches maximum extension (Fig. 1). For the purpose of this discussion, we have assumed that the orbit is circular.

Suppose the first visit yields no detection. The instrument should return for a second attempt after an elapsed time of $t_1$. This is explained as follows. In the first visit, the instrument examined two segments of the orbit, one on either side of the star, and did not observe the planet (Fig. 2a). The optimum time to return for a second visit is when those segments have "rotated" out of view and two new segments, contiguous with the first, have rotated into view (Fig 2b). Return sooner and the instrument will reexamine in part the segments searched during its first visit; return later and the exoplanet may have come and gone in the interim. The time required for first two original segments to rotate out of view is $t_1$.

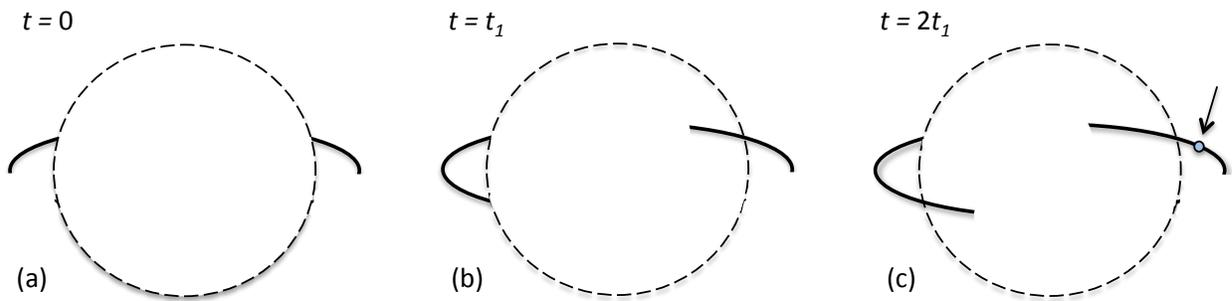

Figure 2. In the first imaging visit (a), the instrument examines two segments of the orbit, one on either side of the star. No planet is observed. In the second visit (b), two new segments, contiguous with the first, are examined. Again no planet is observed. In the third visit (c), two more contiguous segments are examined and the planet is observed. The visits are separated by time $t_1$, the duration of the planet's visibility near maximum extension.

The probability of detection on the second visit has two definitions of operational significance. These are: (i) the probability, $P_2$, of detection after having made two visits; and (ii) the probability, $C_2$, of detection on the second visit after having made no detection on the first visit. The former improves because the region searched has doubled. The latter improves because the region remaining unsearched has become smaller. These two metrics are defined as

$$P_2 = 2P_1 \qquad C_2 = 2\frac{t_1}{T - 2t_1} = \frac{\theta}{\pi - \theta} = \frac{P_1}{1 - P_1} \tag{2}$$

Suppose the second visit yields no detection. The instrument should return for its third visit after another elapsed time of $t_1$ at which time another pair of segments will have rotated into view (Fig. 2c). This process can be continued, with the instrument returning at intervals of $t_1$, until the exoplanet has been imaged.

The number of visits, $N_{max}$, required to examine all segments in the orbit is given by the number required to examine the initial segment AB plus the "front" part of the orbit, segment BC (Fig. 1). Thus,

$$N_{max} = \frac{t_1 + T/2}{t_1} = \frac{\theta + \pi}{\theta} = \frac{1}{P_1} + 1 \tag{3}$$

where the right hand side of eqn. (3) is rounded up to the nearest integer.

In the equations above, the angle $\theta$ is given by (Fig. 3)

$$\cos^2 \theta = \frac{b^2/a^2 - \cos^2 \phi}{\sin^2 \phi} \tag{4}$$

where $\phi$ is the inclination of the orbital plane to the line-of sight from the Earth, $a$ is the angular separation of the planet and star at maximum extension, and $b$ is the instrument IWA. The convention adopted is that for an orbit viewed "edge on," $\phi = 90 \deg$.

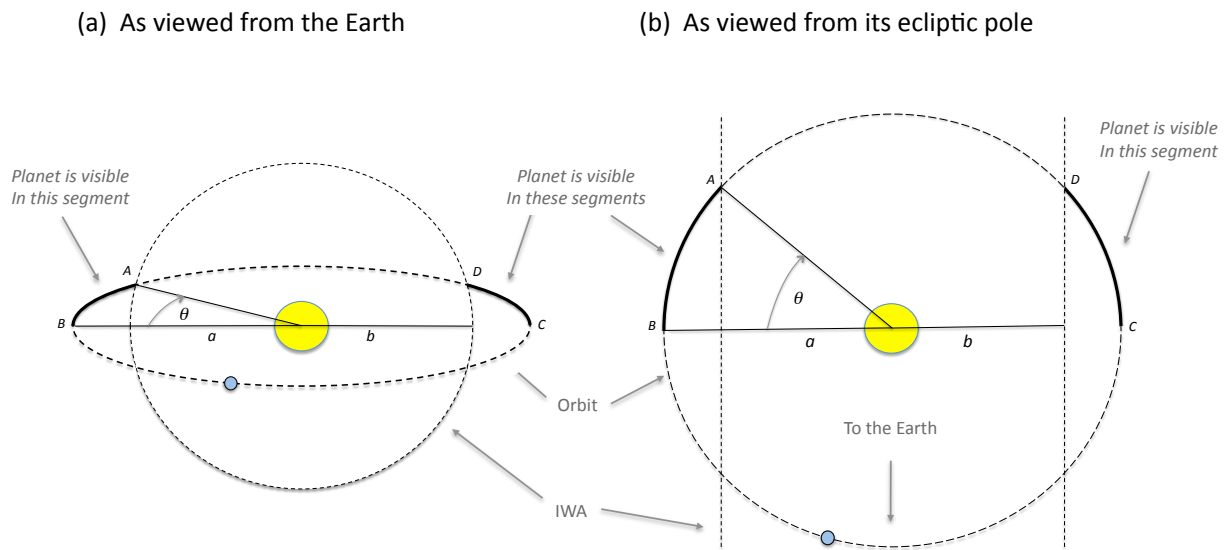

Figure 3. The same system is shown from two perspectives: (a) as viewed from the Earth and (b) as viewed from its own ecliptic pole. If the *IWA = 75 mas*, then the scale of this figure corresponds to a maximum extension of *a = 100 mas* and an orbital inclination of $\phi$ = *75 deg*.

The generalized expression for probability, $P_n$, of detection of the planet after having made $n$ visits carried out according to the prescription above, i.e., with the instrument returning at intervals of $t_1$, is given by:

$$P_n = nP_1 \tag{5}$$

This expression breaks down for the last three visits, i.e., visits $(N_{max}-2)$ to $N_{max}$. There are two reasons for this: (i) the orbit does not contain an integral number of segments of length $\theta$ and (ii) the visible segments are not evenly spaced about the orbit. The modified expressions for these last visits are these:

$$P_{N_{max}-2} = 1 - \frac{(1+r)P_1}{2} \qquad P_{N_{max}-1} = 1 - \frac{rP_1}{2} \qquad P_{N_{max}} \equiv 1 \tag{6}$$

where $r$ is the remainder following the operation

$$r = \frac{1}{P_1} - INT(\frac{1}{P_1}) \tag{7}$$

During each revolution, the planet will make two excursions outside the instrument IWA. Each of these excursions will be of duration $\tau$, where $\tau$ is given by:

$$\tau = \frac{2\theta}{2\pi}T = P_1 T \tag{8}$$

Under the assumption of a quarter-phase threshold for detection, the planet is visible during any one of these excursions for time $t_1$. Thus,

$$t_1 = \frac{P_1 T}{2} \tag{9}$$

Equation (9) gives the interval between visits from the second to the $(N_{max}-1)^{st}$ visits. The interval between the last two visits, if they are necessary, is not fixed but may fall anywhere in the range $rt_1 \rightarrow t_1$. The reason is that the orbit does not contain an integral number of segments of length $\theta$.

The generalized expression for probability $C_n$ of detection of the planet on the $n^{th}$ direct detection visit after having detected no planet on the previous $n-1$ visits is:

$$C_n = \frac{C_1}{1-(n-1)C_1} \tag{10}$$

This expression breaks down for the last three visits. The reasons are the same as in the case of eqn. (5). The modified expressions for these visits are these:

$$C_{N_{max}-2} = \frac{1}{2} \qquad C_{N_{max}-1} = \frac{1}{1+r} \qquad C_{N_{max}} \equiv 1 \qquad (11)$$

Equation (3) represents the *maximum* number of visits required for the detection of the planet. However, it is unlikely that this many will be necessary. And the mission planners would be unlucky in the extreme if this number of visits were required to detect and image every planet known to them from an astrometric precursor.

Thus, it is important to ask what is the *probable* number of visits required for the detection of the planet. Again the result is complicated by the facts that: (i) the orbit does not contain an integral number of segments of length $2\theta$ and (ii) the visible segments are not evenly spaced about the orbit. The *probable* number of visits required is given by the weighted average:

$$N_{prob} = \sum_{n=1}^{N_{max}} w_n n \Big/ \sum_{n=1}^{N_{max}} w_n \qquad (12)$$

where the weights, with the exception of those for the last three visits, are given by $w_n = 1$. The modified expressions for the last three visits are:

$$w_{N_{max}-2} = (1+r)/2 \qquad w_{N_{max}-1} = 1/2 \qquad w_{N_{max}} = r/2 \qquad (15)$$

### Access to the Sky—Some Comments on Feasibility

The implementation of this approach requires that observations of target stars be carried out with a particular cadence. This will have to accommodate, and may be significantly limited by, the imaging instrument's access to the sky (Table 1). Access to the sky is governed by three significant constraints: (i) the instrument's solar exclusion angle(s), (ii) the time to slew between sources, and (iii) the total number of stellar visits allowed in a five-year mission. The primary constraint is that of the solar exclusion angles. All three classes of imaging instruments—external occulters, internal coronagraphs, and visual nullers—have sun exclusion angles, usually quoted in the literature as 45 degrees. (There is one exception—a proposed external occulter mission that utilizes the James Webb Space Telescope (JWST) in conjunction with a starshade (Brown and Soummer 2010; Catanzarite and Shao 2011). JWST has a sun exclusion angle of 85 degrees.) External occulters additionally cannot view targets in the hemisphere opposite the Sun, this being required to prevent contamination of the image from sunlight reflected off the starshade. The starshade exclusion angles as quoted in the literature range from 85 degrees for THEIA (Kasdin et al. 2009) to 105 degrees for the JWST/starshade concept. At any particular epoch, an internal coronagraph and a visual nuller can see about 85% of the sky (Brown et al. 2007; Shao et al. 2008). The THEIA external occulter can see about 31% of the sky. The JWST/starshade occulter can see about 17% of the sky. Targets falling in these exclusion zones cannot be scheduled for observation. This consideration alone for some targets and some instruments will preclude the full implementation of the methodology proposed in this paper.

The second constraint affecting access to the sky is slew rate. Internal coronagraphs and visual nullers move between targets in minutes. But the slew rate for a starshade is about one degree per day. As a consequence, external occulters may require up to 10–14 days to slew between targets. The third major constraint is the number of stellar visits allowed in a five-year mission. Coronagraphs and visual nullers are capable of 300 to 500 visits, limited primarily by the time required to record images and spectra and by the fraction of mission time allocated to planet hunting. For external occulter missions, the number of stellar visits allowed in a five-year mission is about 110 in the case of THEIA (Kasdin et al. 2009; Benson et al. 2009) and about 70 for the proposed JWST/starshade mission (Brown and Soummer 2010). The number of visits in these cases is determined by slew rate and propellant mass.

| Coronagraph (Mission) | Sun Angle (deg) Max | Min | Available Sky* | Slew Time (hr) | Total Visits (5 yr mission) |
|---|---|---|---|---|---|
| External Occulter (JWST+Starshade) | 105 | 85 | 0.17 | ≈250 | ≈70 |
| External Occulter (THEIA) | 85 | 45 | 0.31 | ≈250 | ≈110 |
| Internal Coronagraph (TPF C) | 180 | 45 | 0.85 | ≈1 | 300–500 |
| Visual Nulling Coronagraph (DAViNCI) | 180 | 45 | 0.85 | ≈1 | 300–500 |

* "Available Sky" denotes the fraction of the sky available to the instrument at any given moment.

The most critical of these three constraints is the size of the solar exclusion zone. This determines the instrument's "duty cycle"—the fraction of the time the system is active with respect to any particular target star. The duty cycle is a function of the star's ecliptic latitude and of the coronagraph type. The least impacted parts of the sky are those that lie greater than 45 degrees from the ecliptic (accounting for 30% of the sky.) Stars in these regions are visible at all times for internal coronagraphs and visual nullers and for about six continuous months each year for the THEIA external occulter. Stars located on or very near the ecliptic are the most problematic. These stars are visible for about nine continuous months for internal coronagraphs and visual nullers. For the THEIA external occulter mission, these stars are visible each year during two windows or "seasons," each between six and seven weeks in length; the windows are separated by about 13 weeks. Because of its large sun exclusion zone, stars on the ecliptic would be visible to a JWST/starshade coronagraph during two small windows, each of about three weeks duration. Stars located on the ecliptic are visible each year during two windows, each of about three weeks duration. As this instrument looks to targets removed from the ecliptic, the window increases in length by an operationally inconsequential amount. Figure 4 illustrates the solar exclusion periods for several coronagraphic missions. The imaging instruments cannot view targets when they are in one of the solar exclusion zones. The zones are defined by proximity to the Sun and, in the case of external occulter missions, by the need to avoid contamination of the image from sunlight reflected off the starshade. The duty cycle is indicated on the right side of the figure.

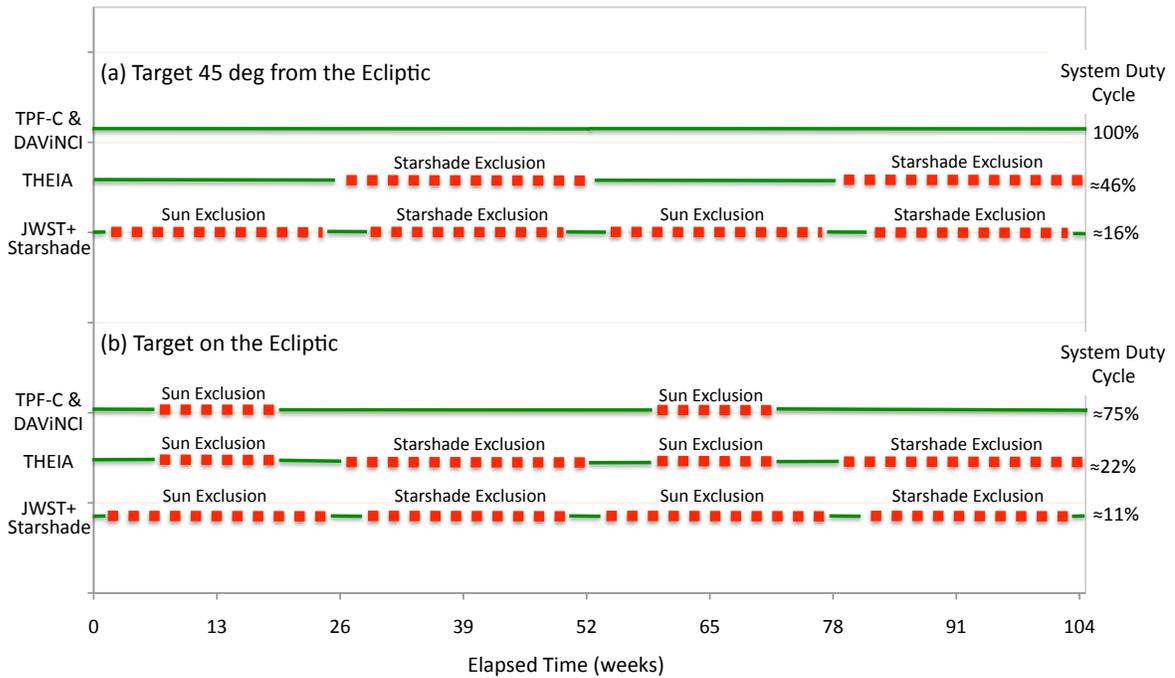

Figure 4. Solar exclusion periods for several coronagraphic missions. Periods of visibility are denoted by solid green lines. The imaging instrument cannot view targets when they are in the exclusion zones (denoted by dashed red lines). The zones are defined by proximity to the Sun or, in the case of external occulter missions, by need to avoid contamination of the image from sunlight reflected off the starshade. The fraction of the time that the system is in an active state (i.e., targets visible to the instrument) is indicated at the right side of the figure.

**Some Examples – Seven Nearby Type F, G, and K Stars**

How would this approach work in practice? We have selected seven F, G, and K stars that appear on the target lists of previous studies (Catanzarite and Shao 2011; Brown and Soummer 2010) of astrometric and imaging mission concepts (Table 2). In addition, we include the Earth-Sun system as it would appear if viewed from a distance of 10 parsecs. This selection is intended to be more-or-less representative of the stars on these lists. In constructing this table, we have made the following assumptions.

- An astrometry mission has surveyed about 100 FGK stars in the solar neighborhood. It has detected terrestrial planets in orbit about 8 of these stars and estimated their masses, orbital parameters, and periods.
  - The planets are in circular orbits at the radius of the "Earth habitable zone," where the EHZ is defined as 1 AU scaled by the square root of the luminosity of the host star.
  - Their orbital inclination is 60 degrees, the most probable, where we have adopted the convention that "edge on" corresponds to an inclination of 90 degrees.
  - The planets all have masses of $1.0\,M_\oplus$.

- A follow-on direct detection mission is charged with recording images and spectra for these planets. The mission planners will use the approach described above to design and carry out an exhaustive search that will capture images for all eight of them. The instrument is designed so that:
  - The planet is visible to the imaging instrument whenever its angular separation from the host star is greater than the instrument IWA and it is in quarter phase or greater.
  - The instrument IWA is 75 mas.

- The questions to be answered are these. Given astrometric orbits:
  - If the first visit yields no detection, when should the instrument return for a second visit? A third visit?
  - What is the first-visit probability of detection for each planet? If the first visit yields no detection, what is the probability after having made two visits? After further visits?
  - What is the *maximum* number of visits required to find each planet and how long will that take if contiguous segments are viewed in sequence?
  - What is the *probable* number of visits required to find each planet. How long will that take?
  - What is the *maximum* number of visits required to find *all* the planets in this pool?
  - What is the *probable* number of visits required to find *all* the planets in this pool?
  - How is the implementation of this idealized approach complicated or limited by operational constraints?

There are seven questions posed in the list above. The answers to the first six of these are summarized in the lower half of Table 2. (The seventh question relates to the feasibility of implementation. This is discussed in detail on a star-by-star basis in later sections.) The upper half of the table summarizes the input data, as it were, including the résumé of the target star (stellar type, distance, mass, luminosity, and stellar reflex semi-amplitude), the résumé of the planet (period, orbital radius, maximum angular extension, duration of visibility, and inclination of the orbit), and the IWA of the instrument. The stellar reflex motion is included to establish that an astrometry mission having a noise floor of 0.035 µas and a 5.8-sigma threshold for detection would detect all of the planets referenced in this table with a probability of ≥99% (Catanzarite et al. 2006; Brown 2009). We now examine the derived results and feasibility of implementation of a star-by-star basis.

Table 2.
Examples taken from the target lists of proposed astrometric* and direct detection** missions.

| Star (Hipparcos ID number): | 73184 | 19849 | 29271 | Sun | 1599 | 105858 | 22449 | 50954 |
|---|---|---|---|---|---|---|---|---|
| **Star's Résumé** | | | | | | | | |
| Stellar type: | K4 | K0/1 | G6 | G2 | G0 | F7 | F6 | F2/3 |
| Distance (pc): | 5.91 | 5.04 | 10.1 | 10 | 8.59 | 9.22 | 8.03 | 16.2 |
| Mass ($M_{Sun}$): | 0.69 | 0.82 | 0.91 | 1.0 | 1.13 | 1.27 | 1.29 | 1.46 |
| Luminosity ($L_{Sun}$): | 0.26 | 0.41 | 0.83 | 1.0 | 1.21 | 1.35 | 2.63 | 4.96 |
| Stellar reflex semi-amplitude ($\mu$as): | 0.38 | 0.46 | 0.30 | 0.31 | 0.34 | 0.30 | 0.47 | 0.28 |
| **Planet's Résumé** | | | | | | | | |
| Period (yr): | 0.44 | 0.56 | 0.91 | 1.0 | 1.09 | 1.11 | 1.82 | 2.75 |
| Radius of orbit (AU): | 0.51 | 0.64 | 0.91 | 1 | 1.10 | 1.16 | 1.62 | 2.23 |
| Maximum angular extension (mas): | 87 | 126 | 90 | 100 | 128 | 126 | 202 | 137 |
| Duration of visibility, $t_1$ (wk) | 2.3 | 5.5 | 5.2 | 7.2 | 11.0 | 11.0 | 47.5 | 29.9 |
| Assumed inclination of orbit (deg): | 60 | 60 | 60 | 60 | 60 | 60 | 60 | 60 |
| **Instrument's Résumé** | | | | | | | | |
| Assumed instrument IWA (mas): | 75 | 75 | 75 | 75 | 75 | 75 | 75 | 75 |
| **Maximum No. Visits** | | | | | | | | |
| Maximum number of visits required: | 7 | 4 | 6 | 5 | 4 | 4 | 2 | 4 |
| Maximum time to find planet (wk): | 11.5 | 14.6 | 23.7 | 26.1 | 28.4 | 29.0 | 47.5 | 71.7 |
| **Probable No. Visits** | | | | | | | | |
| Probable number of visits required: | 3.11 | 2.05 | 2.91 | 2.48 | 2.04 | 2.05 | 1.50 | 1.96 |
| Probable time to find planet (wk): | 4.8 | 5.8 | 10.0 | 10.7 | 11.4 | 11.5 | 23.8 | 28.6 |
| **Interval between Visits** | | | | | | | | |
| Interval between initial visits (wk) | 2.3 | 5.5 | 5.2 | 7.2 | 11.0 | 11.0 | | 29.9 |
| Interval between final two visits (wk) | 0.1 | 3.6 | 2.8 | 4.4 | 6.5 | 7.0 | 47.5 | 11.9 |
| **Probability of detection after n visits, $P_n$** | | | | | | | | |
| Probability of detection first visit, $P_1$: | 0.20 | 0.38 | 0.22 | 0.28 | 0.39 | 0.38 | 0.50 | 0.42 |
| Probability after two visits, $P_2$: | 0.40 | 0.69 | 0.44 | 0.55 | 0.69 | 0.69 | 1.00 | 0.71 |
| Probability after three visits, $P_3$: | 0.60 | 0.88 | 0.66 | 0.78 | 0.88 | 0.88 | | 0.92 |
| Probability after four visits, $P_4$: | 0.80 | 1.00 | 0.83 | 0.92 | 1.00 | 1.00 | | 1.00 |
| Probability after five visits, $P_5$: | 0.90 | | 0.94 | 1.00 | | | | |
| Probability after six visits, $P_6$: | 0.99 | | 1.00 | | | | | |
| Probability after seven visits, $P_7$: | 1.00 | | | | | | | |
| **Additional Derived Results** | | | | | | | | |
| Probability, full orbit is outside IWA: | 0.14 | 0.40 | 0.17 | 0.25 | 0.41 | 0.40 | 0.63 | 0.45 |
| Inclination for orbit outside IWA (deg): | ≤30 | ≤54 | ≤34 | ≤41 | ≤54 | ≤54 | ≤68 | ≤57 |

\* Catanzarite and Shao 2011
\*\* Brown and Soummer 2010

*The Case of Hipparcos 73184*

Hipparcos 73184 is a Type K4 star about 5.91 parsecs from the Earth. A planet orbiting in its "Earth habitable zone," would orbit this star with a period of about 0.44 yr. The star-planet angular separation at maximum extension would be 87 mas. Given an IWA of 75 mas and an orbit inclination of 60 degrees, this planet would be visible to an imaging instrument twice per revolution, each time for about 16 days. The probability of detection in a randomly scheduled first visit is about 20%. If the planet is not detected in the first visit, the instrument would make subsequent visits at intervals of about 16 days until the planet is detected. The probable number of visits required is 3.1, spanning a period of about 34 days. In the worst case, 7 visits would be required, spanning a period of about 80 days.

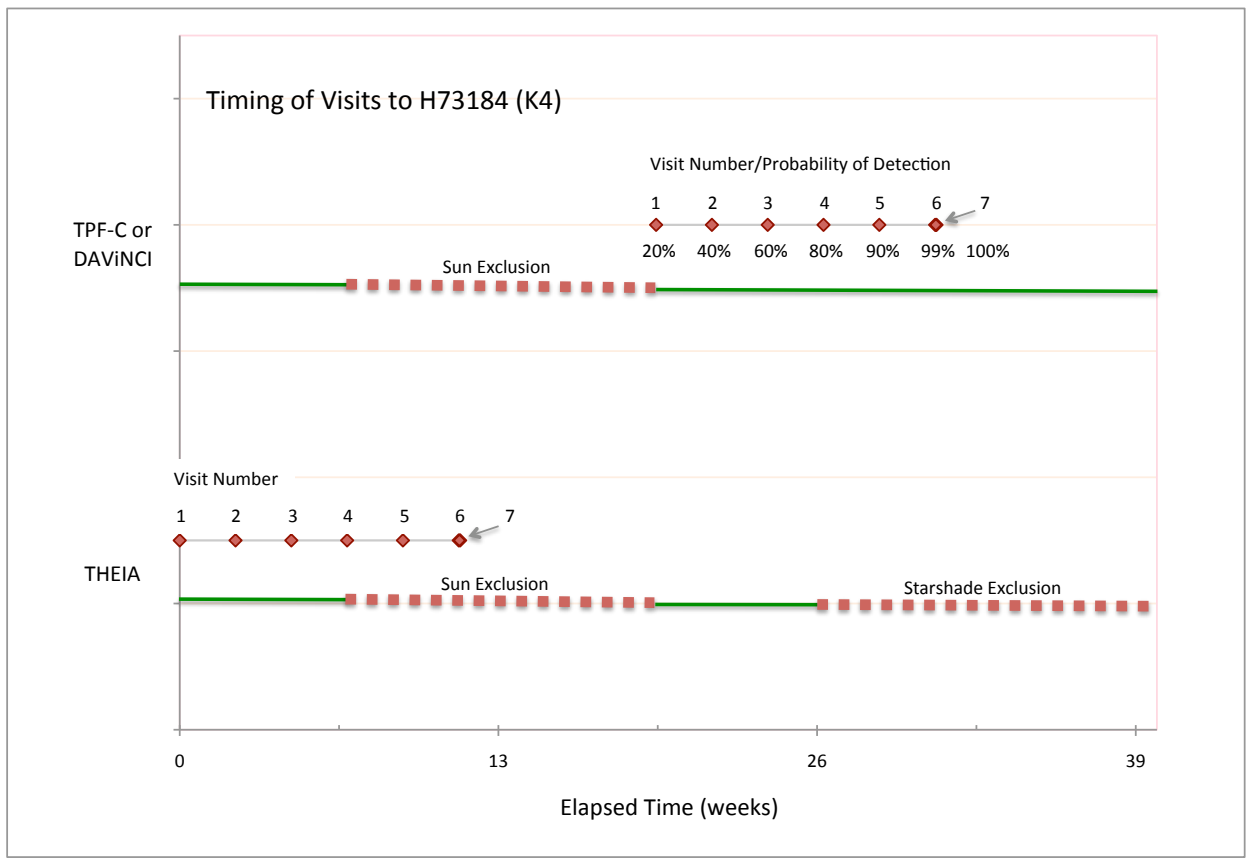

Figure 5. The timing of visits for Hipparcos 73184, a Type K4 star 5.91 parsecs from the Earth. The source is assumed to lie on the Ecliptic, i.e., the worst case. The maximum number of visits required is 7 visits. However, the probable number of visits is only 3.1. The internal coronagraph and the visual nuller mission can easily execute the sequence of seven visits. The external occulter mission can only carry out the first four visits, after which the window of visibility closes. However, the probability of having detected the planet after four visits is 79.6%. If further visits are needed, they can be scheduled during future windows of visibility.

This progression is illustrated in Fig. 5 for the three coronagraph types. The source is assumed to lie on the Ecliptic, i.e., the worst case.  As this figure shows, the internal coronagraph and the visual nuller mission can easily execute the sequence of seven visits, scheduling the first visit shortly after the star has emerged from the solar exclusion zone.  The external occulter mission can only carry out the first four visits, after which its window of visibility closes.  The occulter may still benefit from the application of this approach.  First, the probability of its having detected the planet after those four visits is fairly high, about 79.6%.  Thus, it is probable that visits number 5, 6, and 7 would not be needed.  Second, the segments need not be observed in strict chronological sequence.  If an operational constraint intrudes, e.g., the target star is inside the solar or starshade exclusion zones at the time of a scheduled observation, then the observation of the "missed segments" can be scheduled to occur one or more revolutions of the exoplanet in the future when the appearance of these segments aligns with the occulter's window of visibility.  Third, we note that if this star did not lie on the Ecliptic, but lay ≥ 45 degrees from it, the length of the Sun exclusion window would go to zero and the external occulter could easily execute the full sequence of seven visits.

*The Case of Hipparcos 19849*

Hipparcos 19849 is a Type K0/1 star about 5.04 parsecs from the Earth.  A planet orbiting in its "Earth habitable zone," would orbit this star with a period of about 0.56 yr.  The star-planet angular separation at maximum extension would be 126 mas.  Given an IWA of 75 mas and an orbit inclination of 60 degrees, this planet would be visible to an imaging instrument twice per revolution, each time for about 38 days.  The probability of detection in a randomly scheduled first visit is about 38%.  If the planet is not detected in the first visit, the instrument would make subsequent visits at intervals of about 38 days until the planet is detected. The probable number of visits required is 2.0, spanning a period of about 40 days.  In the worst case, 4 visits would be required, spanning a period of about 24 weeks.

This progression is illustrated in Fig. 6 for the three coronagraph types. The source is assumed to lie on the Ecliptic, i.e., the worst case.  As this figure shows, the internal coronagraph and the visual nuller mission can easily execute the sequence of four visits, scheduling the first visit shortly after the star has emerged from the solar exclusion zone.  The external occulter mission can only carry out the first two visits, after which its window of visibility closes.  The occulter may still benefit from the application of this approach.  First, the probability of its having detected the planet after those two visits is fairly high, about 68.9%.  Thus, it is probable that visits number 3 and 4 would not be needed.  Second, the segments need not be observed in strict chronological sequence.  If an operational constraint intrudes, e.g., the target star is inside the solar or starshade exclusion zones at the time of a scheduled observation, then the observation of the "missed segments" can be scheduled to occur one or more revolutions of the exoplanet in the future when the appearance of these segments aligns with the occulter's window of visibility.  Third, we note that if this star did not lie on the Ecliptic, but lay ≥ 45 degrees from it, the length of the Sun exclusion window would go to zero and the external occulter could easily execute the full sequence of seven visits.

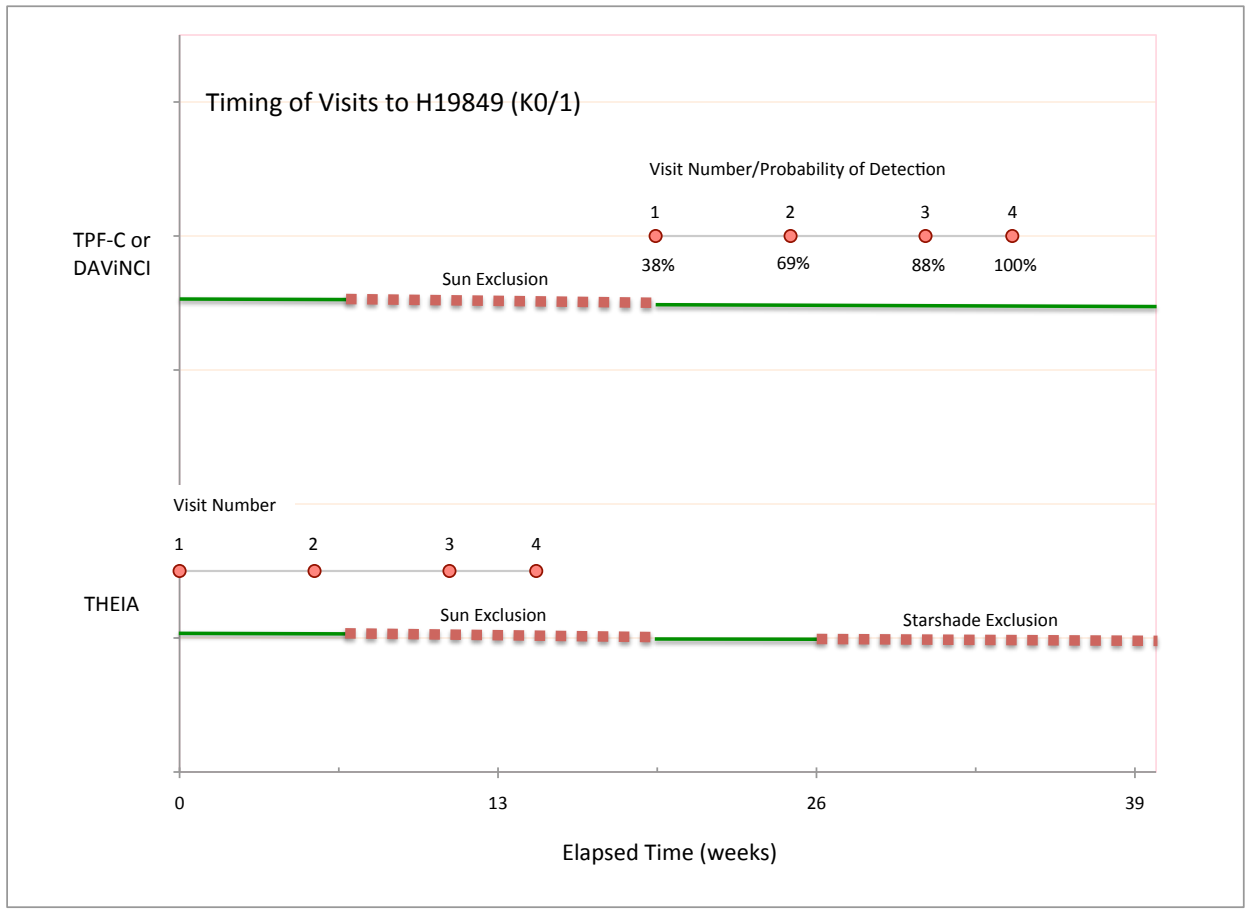

Figure 6. The timing of visits for Hipparcos 19849, a Type K0/1 star 5.04 parsecs from the Earth. The source is assumed to lie on the Ecliptic, i.e., the worst case. The maximum number of visits required is 4 visits. However, the probable number of visits is only 2.0. The internal coronagraph and the visual nuller mission can easily execute the sequence of four visits. The external occulter mission can only carry out the first two visits, after which the window of visibility closes. However, the probability of having detected the planet after two visits is 68.9. If further visits are needed, they can be scheduled during future windows of visibility.

*The Case of Hipparcos 29271*

Hipparcos 29271 is a Type G6 star about 10.1 parsecs from the Earth. A planet orbiting in its "Earth habitable zone," would orbit this star with a period of about 0.91 yr. The star-planet angular separation at maximum extension would be 90 mas. Given an IWA of 75 mas and an orbit inclination of 60 degrees, this planet would be visible to an imaging instrument twice per revolution, each time for about 36 days. The probability of detection in a randomly scheduled first visit is about 22%. If the planet is not detected in the first visit, the instrument would make subsequent visits at intervals of about 36 days until the planet is detected. The probable number of visits required is 2.9, spanning a period of about 70 days. In the worst case, 6 visits would be required, spanning a period of almost 24 weeks.

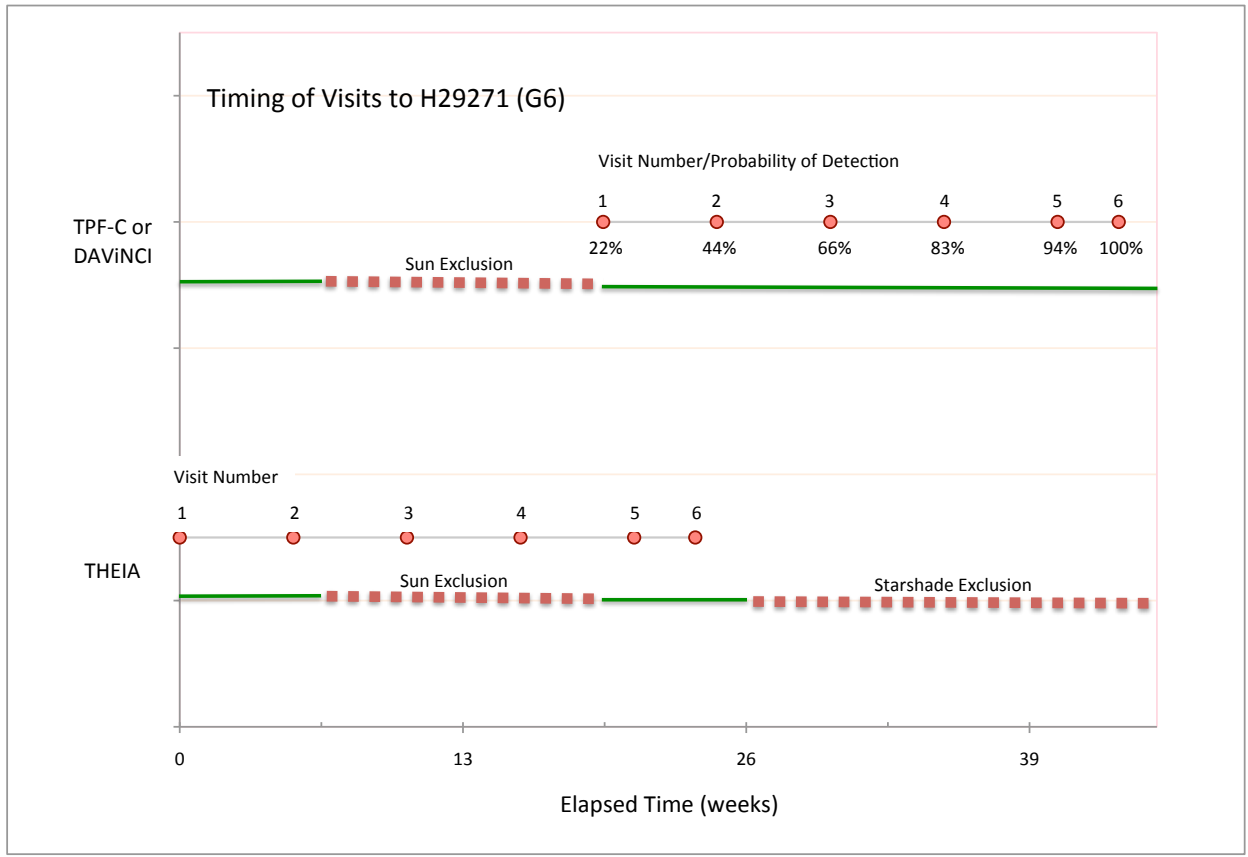

Figure 7. The timing of visits for Hipparcos 29271, a Type G6 star 10.1 parsecs from the Earth. The source is assumed to lie on the Ecliptic, i.e., the worst case. The maximum number of visits required is 6 visits. However, the probable number of visits is only 2.9. The internal coronagraph and the visual nuller mission can easily execute the sequence of seven visits. The external occulter mission can only carry out the first two and the final two visits. However, the probability of having detected the planet after these four visits is about 61%. If further visits are needed, they can be scheduled during future windows of visibility.

This progression is illustrated in Fig. 7 for the three coronagraph types. The source is assumed to lie on the Ecliptic, i.e., the worst case.  As this figure shows, the internal coronagraph and the visual nuller mission can easily execute the sequence of seven visits.  On the other hand, external occulter mission can only carry out the first two and the last two visits.  However, the probability of its having detected the planet after those four visits is fairly high, about 67%. If further visits were needed, they could be scheduled during the occulter's future windows of visibility.  We note that if this star did not lie on the Ecliptic, but lay ≥ 45 degrees from it, the length of the Sun exclusion window would go to zero and the external occulter could easily execute the full sequence of six visits.

*The Case of the Sun (viewed from 10 pc)*

It is instructive to ask how the Earth-Sun system would appear if viewed from 10 parsecs. The Sun is a Type G2 star. The Earth orbits the Sun at a distance of 1 AU with a period of 1 yr. The star-planet angular separation at maximum extension would be 100 mas. The Earth would be visible to the imaging instrument twice per revolution, each time for about 50 days. The probability of detection in a randomly scheduled first visit would be about 28%. If the Earth were not detected in the first visit, the instrument would make subsequent visits at intervals of about 50 days until the planet was detected. The probable number of visits required would be 2.48, spanning a period of about 75 days. In the worst case, 5 visits would be required, spanning a period of about 26 weeks.

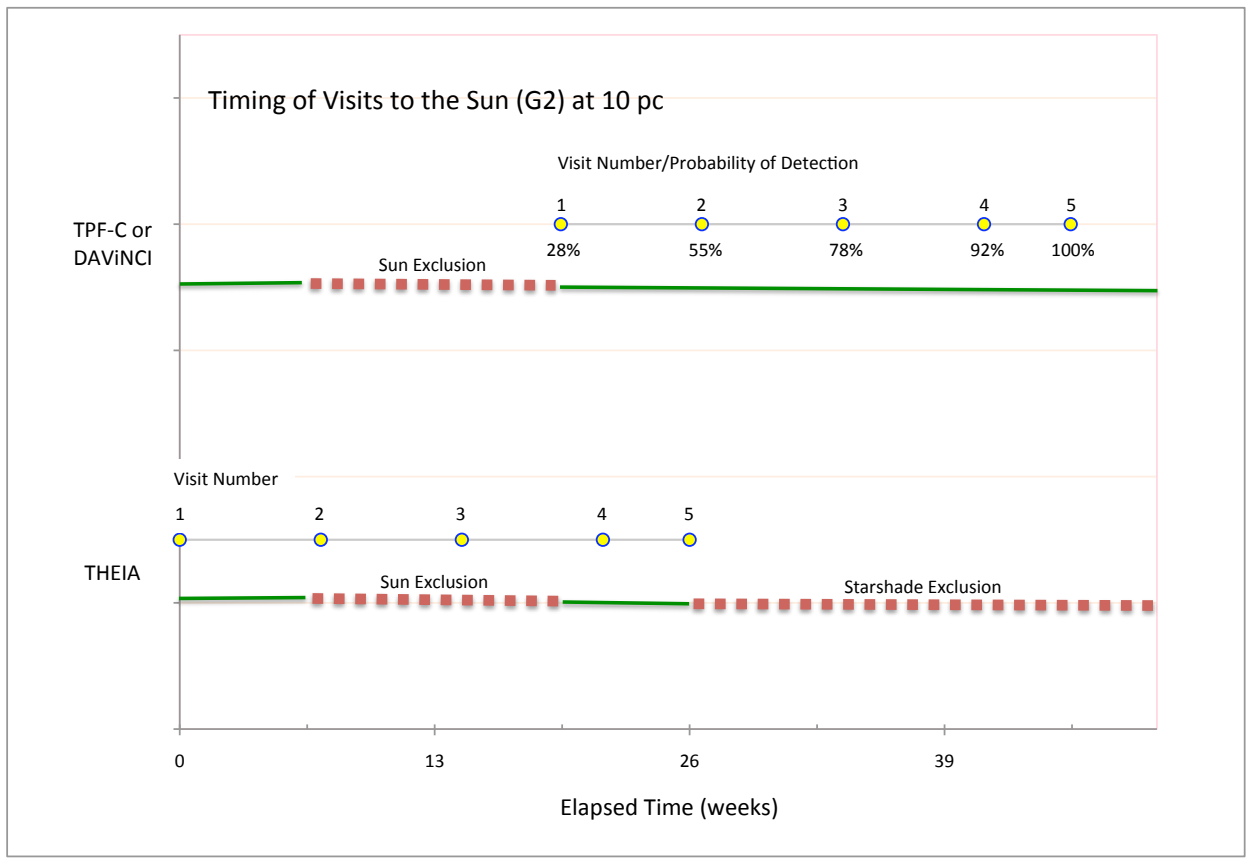

Figure 8. The timing of visits for the Sun-Earth system as viewed from 10 parsecs. The source is assumed to lie on the Ecliptic, i.e., the worst case. The maximum number of visits required is 5 visits. However, the probable number of visits required is only 2.48. The internal coronagraph and the visual nuller mission can easily execute the sequence of five visits. The external occulter mission will miss the third visit. However, the probability of having detected the planet after having made all other visits is about 78%.

This progression is illustrated in Fig. 8 for the three coronagraph types. The source is assumed to lie on the Ecliptic, i.e., the worst case. As this figure shows, the internal coronagraph and the visual nuller mission can easily execute the sequence of five visits. On the other hand, external occulter mission will miss the third visit. However, the probability of its having detected the planet after having made all other visits is high, about 78%. We note that if this star did not lie on the Ecliptic, but lay ≥ 45 degrees from it, the length of the Sun exclusion window would go to zero and the external occulter could execute all visits.

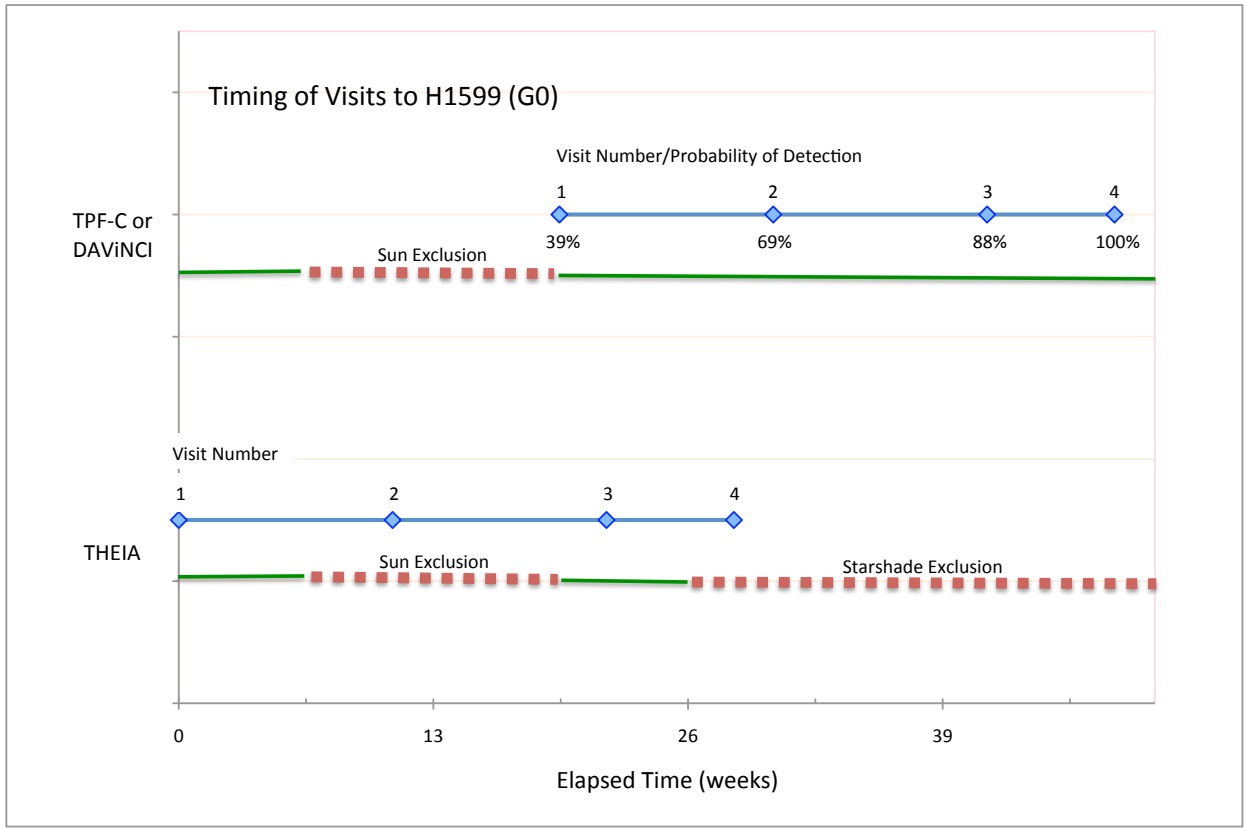

Figure 9. The timing of visits for Hipparcos 1599, a Type G0 star 8.6 parsecs from the Earth. The source is assumed to lie on the Ecliptic, i.e., the worst case. The maximum number of visits required is 4 visits. However, the probable number of visits is only 2.04. The internal coronagraph and the visual nuller mission can easily execute the sequence of four visits. The external occulter mission can only carry out the first and third visits. However, the probability of having detected the planet after these two visits is about 58%.

*The Case of Hipparcos 1599*

Hipparcos 1599 is a Type G0 star about 8.59 parsecs from the Earth. A planet orbiting in its "Earth habitable zone," would orbit this star with a period of about 1.09 yr. The star-planet angular separation at maximum extension would be 128 mas. Given an IWA of 75 mas and an orbit inclination of 60 degrees, this planet would be visible to an imaging instrument twice per revolution, each time for about 77 days. The probability of detection in a randomly scheduled first visit is about 39%. If the planet is not detected in the first visit, the instrument would make subsequent visits at intervals of about 77 days until the planet is detected. The probable number of visits required is 2.04, spanning a period of about 80 days.  In the worst case, 4 visits would be required, spanning a period of about 28 weeks.

This progression is illustrated in Fig. 9 for the three coronagraph types. The source is assumed to lie on the Ecliptic, i.e., the worst case.  As this figure shows, the internal coronagraph and the visual nuller mission can easily execute the sequence of four visits.  On the other hand, external occulter mission can only carry out the first and the third visits.  However, the probability of its having detected the planet after those two visits is about 58%.  If further visits were needed, they could be scheduled during the occulter's future windows of visibility.  We note that if this star did not lie on the Ecliptic, but lay ≥ 45 degrees from it, the length of the Sun exclusion window would go to zero and the external occulter could also execute the third of four visits, increasing the probability of detection to 88.5%.

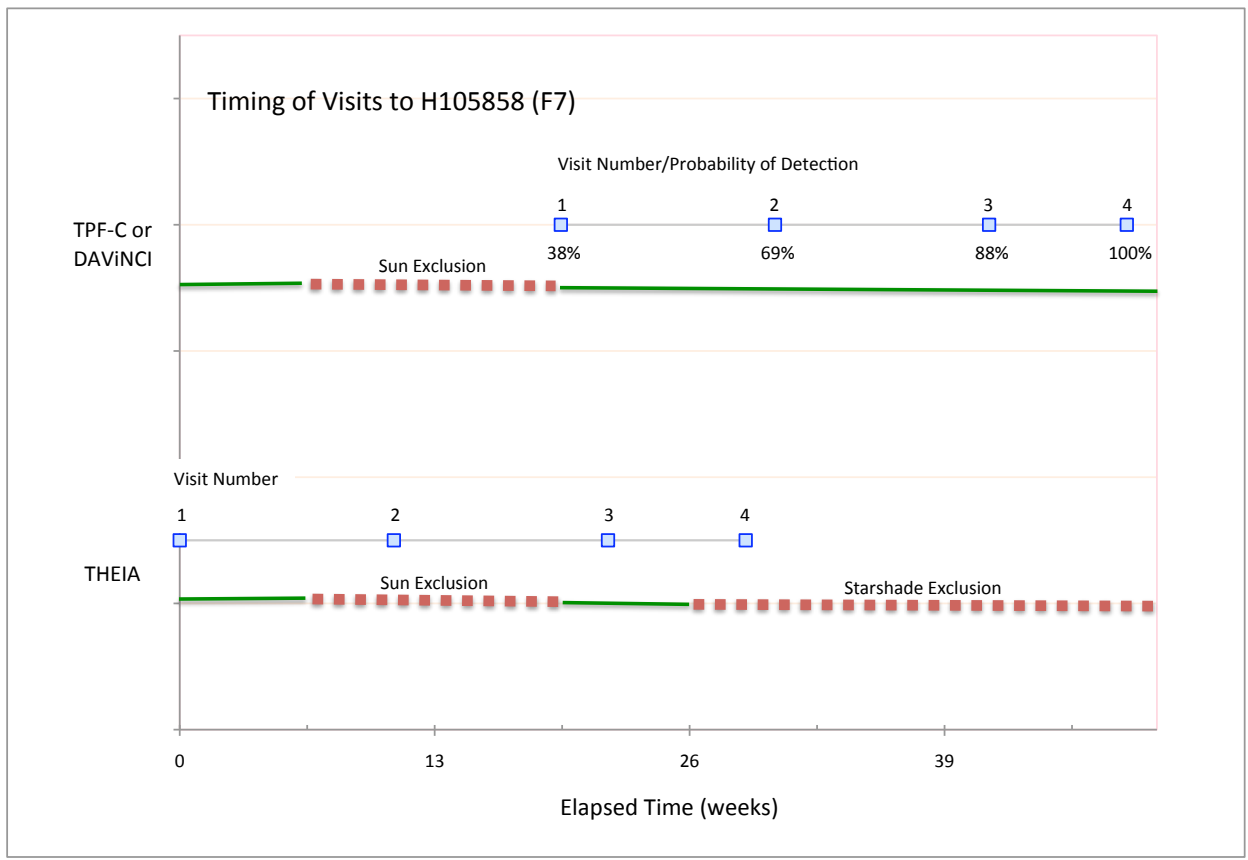

Figure 10.  The timing of visits for Hipparcos 105858, a Type F7 star 9.2 parsecs from the Earth.  The source is assumed to lie on the Ecliptic, i.e., the worst case.  The maximum number of visits required is 4 visits.  However, the probable number of visits is only 2.05. The internal coronagraph and the visual nuller mission can easily execute the sequence of four visits.  The external occulter mission can only carry out the first and third visits.  However, the probability of having detected the planet after having made these two visits is about 57%.

*The Case of Hipparcos 105858*

Hipparcos 105858 is a Type F7 star about 9.22 parsecs from the Earth. A planet orbiting in its "Earth habitable zone," would orbit this star with a period of about 1.11 yr. The star-planet angular separation at maximum extension would be 126 mas. Given an IWA of 75 mas and an orbit inclination of 60 degrees, this planet would be visible to an imaging instrument twice per revolution, each time for about 77 days. The probability of detection in a randomly scheduled first visit is about 38%. If the planet is not detected in the first visit, the instrument would make subsequent visits at intervals of about 77 days until the planet is detected. The probable number of visits required is 2.05, spanning a period of about 80 days. In the worst case, 4 visits would be required, spanning a period of about 29 weeks.

This progression is illustrated in Fig. 10 for the three coronagraph types. The source is assumed to lie on the Ecliptic, i.e., the worst case. As this figure shows, the internal coronagraph and the visual nuller mission can easily execute the sequence of four visits. On the other hand, external occulter mission can only carry out the first and third visits. However, the probability of its having detected the planet after having made those two visits is moderately high, about 57%. We note that if this star did not lie on the Ecliptic, but lay ≥ 45 degrees from it, the length of the Sun exclusion window would go to zero and the external occulter could execute the first three visits, with the probability of detection increasing to about 88%.

*The Case of Hipparcos 22449*

Hipparcos 22449 is a Type F6 star about 8.03 parsecs from the Earth. A planet orbiting in its "Earth habitable zone," would orbit this star with a period of about 1.82 yr. The star-planet angular separation at maximum extension would be 202 mas. Given an IWA of 75 mas and an orbit inclination of 60 degrees, the entire orbit of this planet would lie outside the IWA. However, the planet will be at quarter phase or greater only half of its year, a period lasting about 48 weeks. The probability of detection in a randomly scheduled first visit is 50%. If the planet is not detected in the first visit, the instrument would make a second visit 48 weeks later, at which time the planet would be detected. The probable number of visits required is 1.5, spanning a period of about 24 weeks. In the worst case, 2 visits would be required, spanning a period of about 48 weeks.

This progression is illustrated in Fig. 11 for the three coronagraph types. The source is assumed to lie on the Ecliptic, i.e., the worst case. As this figure shows, all three coronagraphs can execute both visits per the nominal timing.

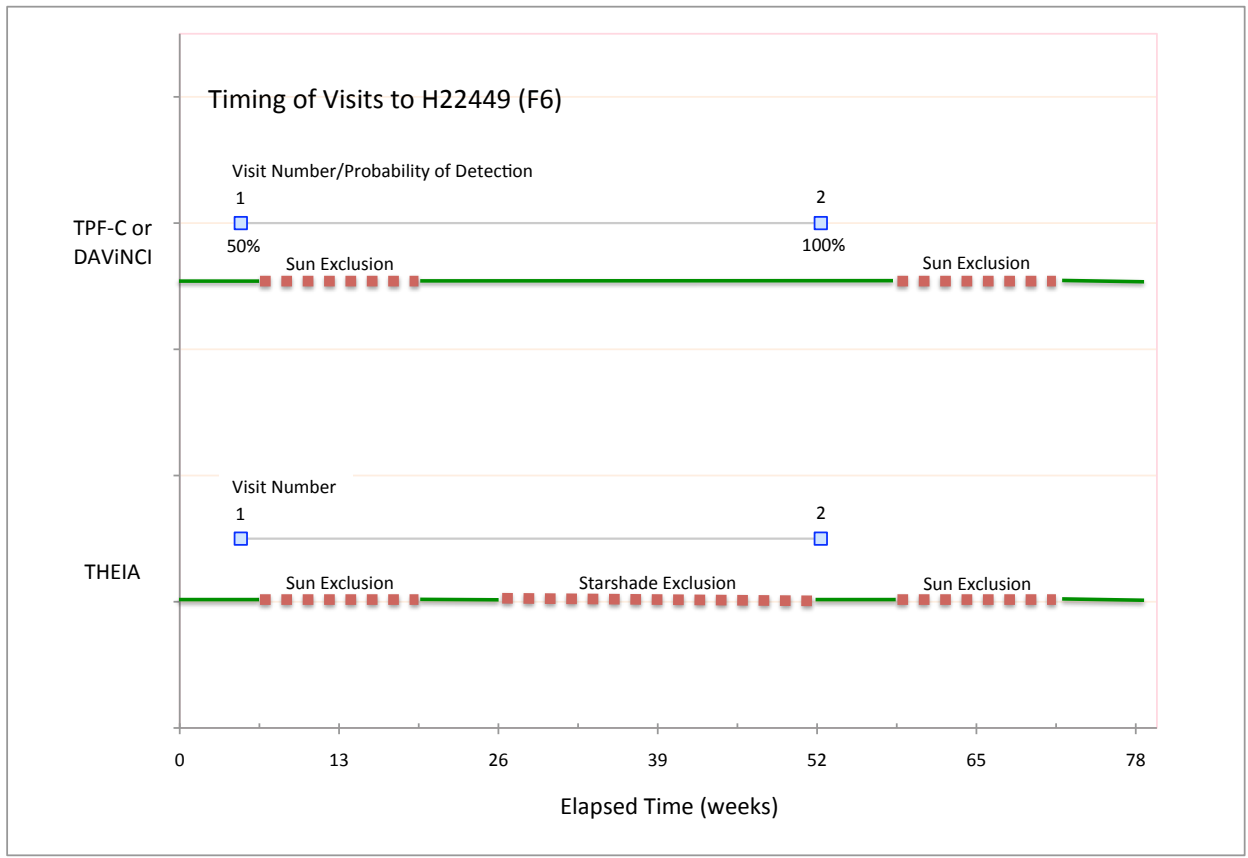

Figure 11. The timing of visits for Hipparcos 22449, a Type F6 star 8.03 parsecs from the Earth. The source is assumed to lie on the Ecliptic, i.e., the worst case. The maximum number of visits required is 2 visits. However, the probable number of visits is only 1.5. All coronagraph types can execute both visits at the appropriate times.

*The Case of Hipparcos 50954*

Hipparcos 50954 is a Type F2/3 star about 16.2 parsecs from the Earth. A planet orbiting in its "Earth habitable zone," would orbit this star with a period of about 2.75 yr. The star-planet angular separation at maximum extension would be 137 mas. Given an IWA of 75 mas and an orbit inclination of 60 degrees, this planet would be visible to an imaging instrument twice per revolution, each time for about 210 days or 30 weeks. The probability of detection in a randomly scheduled first visit is about 42%. If the planet is not detected in the first visit, the instrument would make subsequent visits at intervals of about 30 weeks until the planet is detected. The probable number of visits required is 1.96, spanning a period of about 29 weeks. In the worst case, 4 visits would be required, spanning a period of about 1.4 years.

This progression is illustrated in Fig. 12 for the three coronagraph types. The source is assumed to lie on the Ecliptic, i.e., the worst case. As this figure shows, none of the three coronagraph types can execute the "ideal" observation sequence. However, a slightly modified sequence will permit the full recovery of this loss. This is discussed in the next section..

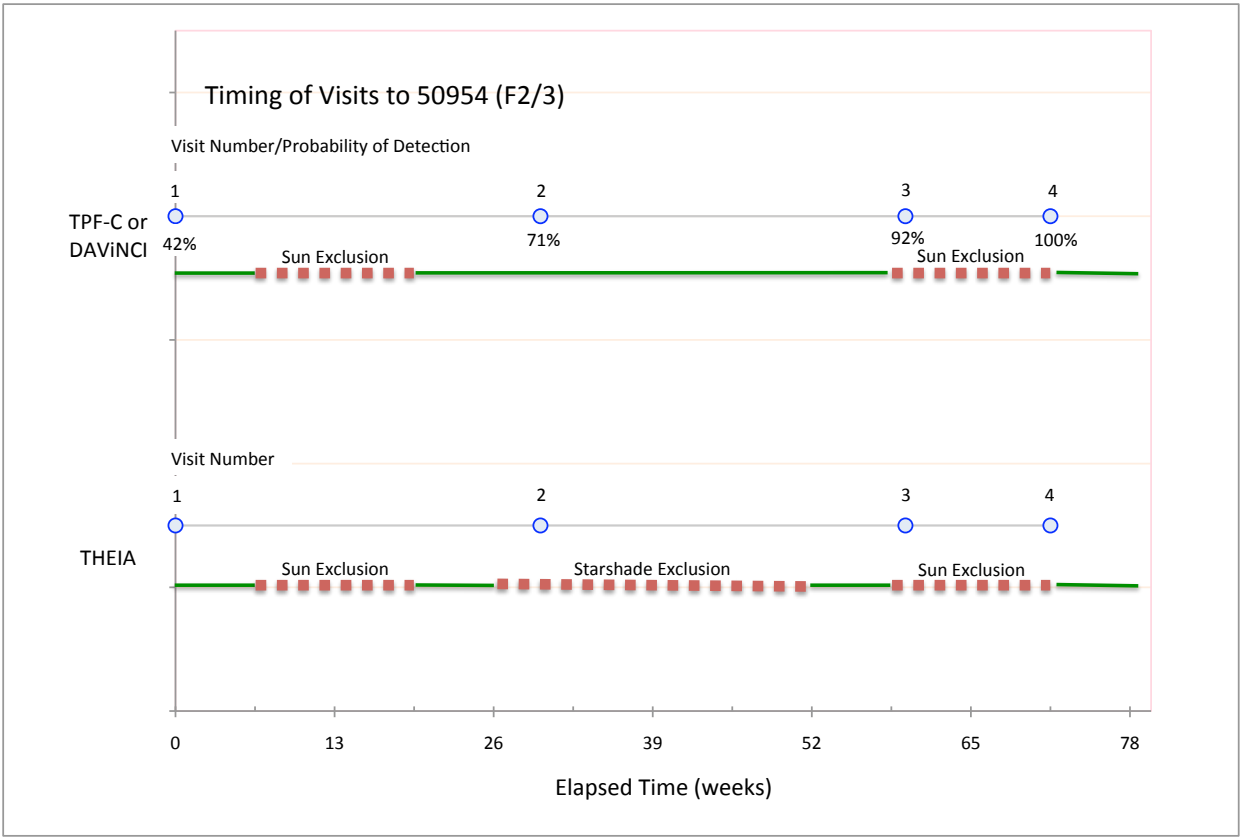

Figure 12. The timing of visits for Hipparcos 50954, a Type F2/3 star 16.2 parsecs from the Earth. The source is assumed to lie on the Ecliptic, i.e., the worst case. The maximum number of visits required is 4 visits. However, the probable number of visits is only 1.96. None of the coronagraph types can execute the "ideal" sequence. A modified sequence, discussed in the next section, permits a full recovery of the lost visits and 100% probability for the detection of the planet.

**Reclaiming the Sky—Further Comments on Feasibility**

The implementation of the "ideal" observing sequence (i.e., one having a 100% probability of detection) may not be possible for some targets and some coronagraph types. However, it is still possible to use the methodology described in this paper to plan and execute a "best available" sequence of observations that will maximize the chances of—if not guarantee—the detection of the planet. A few examples will suffice to illustrate.

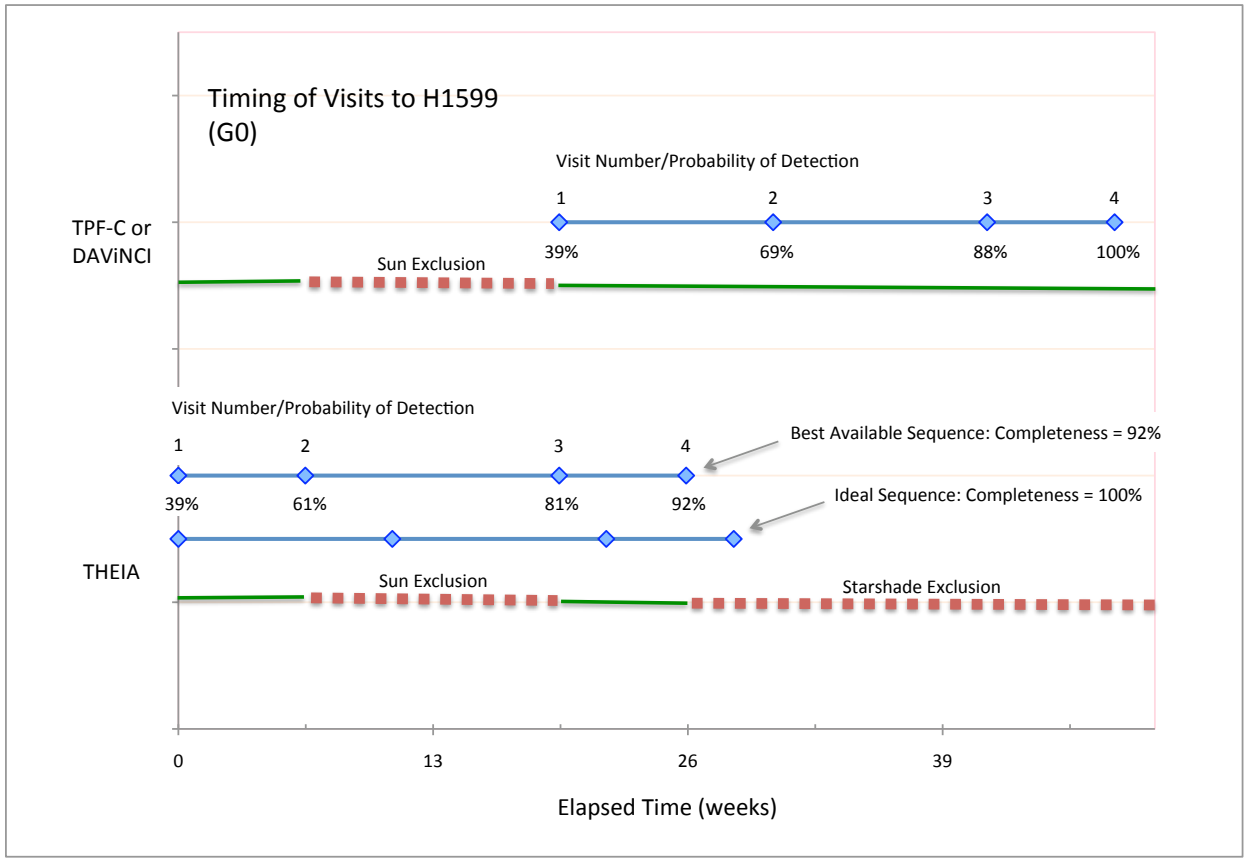

Figure 13. The timing of visits for Hipparcos 1599 is shown. The "ideal" sequence of visits cannot be executed for the THEIA external occulter mission. However, the "best available" sequence provides a probability of detection of 92%.

Consider the case of Hipparcos 1599. Full implementation of our approach calls for a time-independent first visit followed by additional visits, if needed, at 11 weeks, 22 weeks, and 28 weeks. This sequence cannot be carried out for the THEIA external occulter mission, because some of the visits occur when the target is within the sun or starshade exclusion zones. (See Fig. 9.) In this case, a modified sequence can be used that maximizes the likelihood of detection, while conforming to operational constraints. For Hipparcos 1599, this sequence begins as before with a visit that coincides with the appearance of the target in the occulter's first window of visibility. This is followed by additional visits, if needed, at 6.5 weeks, 19.5 weeks, and 26 weeks. This sequence of four visits leads to a probability of detection of 92%. It is illustrated in Fig. 13. This figure is identical to Fig. 9 except that the "best available" sequence has been overlaid.

Consider next the case of Hipparcos 50954. None of the coronagraph types are capable of executing the ideal sequence. (See Fig. 12.) However, there is an alternative sequence of four visits that conforms to operational constraints and still results in a probability of detection of 100%. The reason that this is possible is that the more generalized requirement the separation of any two observations must be less than the planet's duration of visibility, $t_1$, (eqn. 11). As long as this condition is imposed, the instrument will reexamine in part the segments viewed during its previous visit. But there is no risk that the planet will have come and gone in the interim. The alternative sequence is illustrated in Fig. 14.

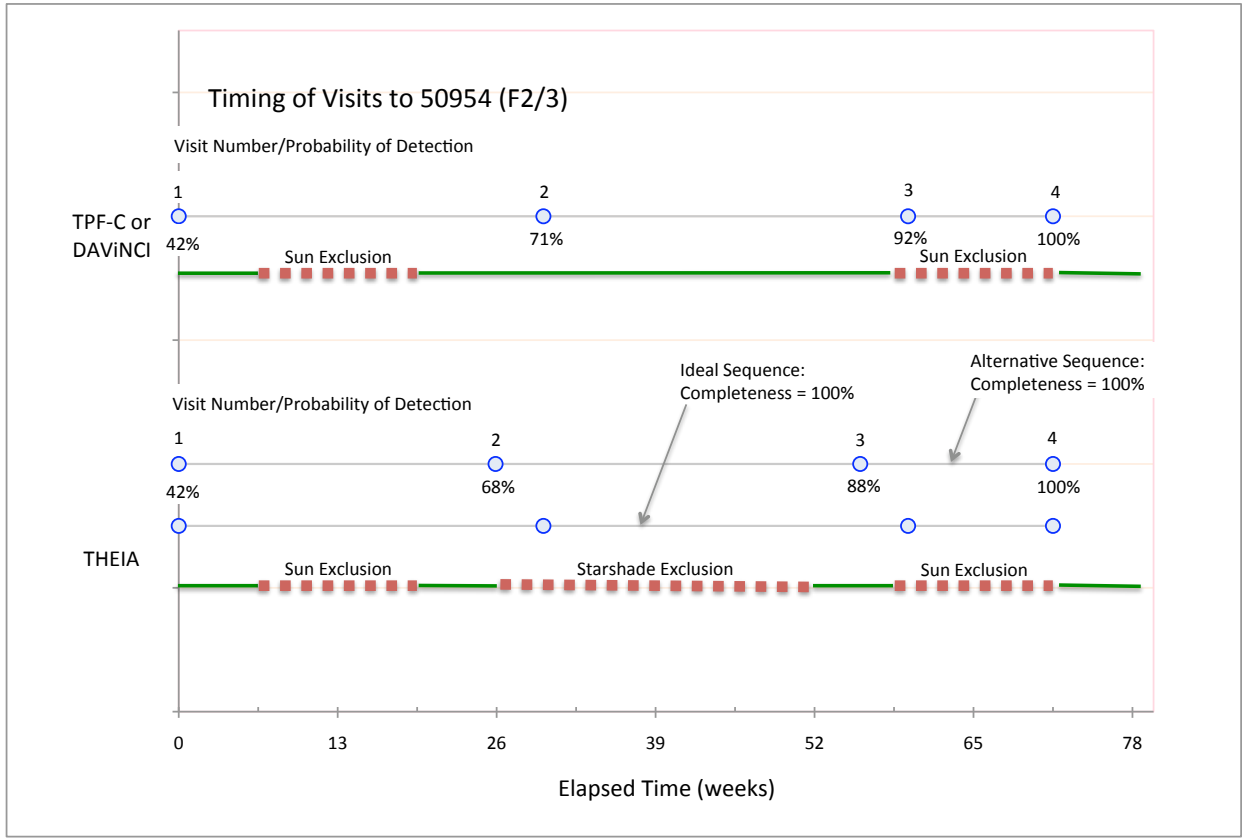

Figure 14. The timing of visits for Hipparcos 50954 is shown. The "ideal" sequence of visits cannot be executed. However, an alternative sequence is available that still achieves a probability of detection of 100% for all three of the coronagraph types.

**Retrospective and Commentary on the Report of the Exoplanet Task Force**

The report of the Exoplanet Task Force (Lunine, J. I. 2008) called for a comprehensive program to detect and characterize terrestrial planets in the solar neighborhood. Five criteria were cited as necessary for an exo-Earth's existence to be established: (i) detection, (ii) confirmation, (iii) mass estimation, (iv) orbit estimation, and (v) the recording of atmospheric spectra. No single mission is capable of meeting all these criteria.

The Task Force recommended a two-mission program. An astrometry mission would fly first and be followed by a direct detection mission. The astrometry mission would perform detection, mass estimation, and orbit estimation. The direct detection mission would confirm the original detections and record atmospheric spectra. This was a sensible ordering. Astrometry is vastly superior at detection and orbit estimation but cannot obtain spectra; direct detection is uniquely capable of obtaining spectra but cannot estimate mass (e.g., Catanzarite and Shao 2011). The maturity of astrometric technology played well to this ordering.

The Task Force made a key assumption: that the imaging mission—inherently weak at detection—would get a strong helping hand from the astrometry mission. Given astrometric orbits, the imager would be able to plan and schedule its visits to coincide with the periods, possibly of short duration, when the exoplanet was visible. Imaging completeness, typically ≈35%, would be increased to ≈100%. The two-mission program would deliver full characterization for all terrestrial planets in the solar neighborhood (Shao et al. 2010).

This assumption was called into question by two studies, showing that orbital phase knowledge degrades rapidly with time (Savransky et al. 2009; Brown 2009.) By the time the direct detection mission was flown, five years or more after the astrometry mission, it would no longer be possible to use astrometric orbits to schedule imaging visits. One of these studies showed that if the THEIA external occulter mission pared its target list to include only those stars known to host terrestrial planets, it would approximately double the detection rate for $\eta_\oplus \leq 0.3$.[3] But this was judged by the authors to be a minor benefit.

Faced with these developments, the exoplanet community divided itself into two camps. One maintained that the five criteria of the Exoplanet Task Force remain the standard upon which a program should be based. Two missions should be launched to identify and fully characterize terrestrial planets in the solar neighborhood. The second camp advocated a program with narrower goals. A single mission, an imaging mission, should be launched to detect and partially characterize terrestrial planets in the solar neighborhood. It would not determine their masses, and in most cases it would not determine their orbits. It would detect only about a third[4] of them. It would record spectra for fewer still. The relative merits of these approaches—comprehensive vs. limited, systematic vs. quick, expensive vs. cheap—continues as a topic of discussion within the community.

Meanwhile, the linkage between astrometric discovery and imaging detection remained broken. Some came to believe that it could not be repaired, that imaging missions are inherently incapable of using astrometric orbits to improve their capacity for detection. It is largely this belief that has sustained the debate over the future of the exoplanet program.

---

[3] *$\eta_\oplus$ (pronounced "eta-Earth") denotes the fraction of stars that host habitable zone terrestrial planets.*

[4] *The completeness for detection varies between different imaging mission concepts. It has a significant dependence on the makeup of the mission target list. Published simulations for the THEIA external occulter mission forecast completeness of about 35%.*

## Discussion and Conclusion

We have shown that astrometric orbits, properly applied, can increase the yield of images and spectra from a follow-on direct detection mission. We have applied this approach to a sample of seven stars, taken from astrometry and imaging target lists, and to the sun-earth system as it would appear from a distance of 10 parsecs. These eight stars were each assumed to have an earth-mass planet orbiting at 1 AU scaled to luminosity. We assumed that astrometric observations had resulted in the detection of all eight planets and led to estimates of their three-dimensional orbit parameters and periods. We found that a direct detection mission using our approach would record images for all 8 planets in this pool with a probability of 100% and that it would require in the worst case a *maximum* of 36 imaging visits. The *probable* number of visits would be considerably smaller, about 20. We examined how implementation of our new approach is complicated and limited by operational constraints. We found that it could be applied to internal coronagraph and visual nuller missions with a success rate approaching 100%. External occulter missions will also benefit, but to a lesser degree.

This is a significant improvement over the performance of a direct detection mission flying in the absence of prior information. Absent prior information, the allocation of stellar visits will be weighted heavily in favor of detection at the expense of characterization. Savransky et al. (2010) have shown, for example, that an external occulter mission would detect on average about 35% of the planets in its target pool, i.e., only 3 of the 8 planets in our sample, and that it would record spectra for only one of them. It would provide no mass estimates and likely no orbit estimates. Moreover, it would require over 100 stellar visits and the full five years of the mission lifetime to do so. If $\eta_\oplus \leq 0.1$, as mounting evidence suggests (Catanzarite and Shao 2011), then the number of exo-Earths in the solar neighborhood (and therefore accessible to an imaging mission) could be 10 or less. Thus, the sample of eight systems examined in this work may be representative of the pool that has been provided to us in nature. Such a rarity of exo-Earths would have important implications. In particular, it would raise the possibility that an imaging mission flying blind would detect zero exo-Earths (Catanzarite and Shao 2011).

The situation is quite different if there exists a methodology that guarantees imaging detection given astrometric discovery. Given prior information, an imaging mission's allocation of visits could be dedicated almost entirely to characterization. For 8 systems in our sample, only 20 visits on average would be required for reacquisition of all 8 planets. The remaining visits (about 90 in the case of an external occulter mission) would be available for recording spectra and searching, for example, for diurnal variations indicative of topography and geography ("oceans and continents"), longer scale variations indicative of weather, and annual variations indicative of seasons. Moreover, the overall program would deliver mass and orbit estimates for all exoplanets.

In summary, astrometric or coronagraphic missions, taken separately, would deliver limited results. An astrometric mission would deliver partial characterization for all planets in the solar neighborhood (Traub et al. 2008). This characterization would not include atmospheric spectra. A coronagraphic mission would deliver partial characterization for about a third of the planets in the solar neighborhood (Savransky et al. 2010).  This characterization would provide no mass estimates, likely no orbit estimates, and the yield could include false positive detections (Catanzarite and Shao 2011).  But in combination, the two mission types would deliver full and accurate characterization for all exo-Earths in the solar neighborhood.  Given the possibility that this number is small—10 or fewer, the combined approach described in this paper deserves consideration in planning the way forward for exoplanet research.

The analysis presented in this paper does not incorporate all effects of interest and importance. In particular, it does not address the effects of standard errors and non-circular orbits. Moreover, the semi-major axes and three-dimensional orientations of the orbits were not selected at random and the author selected the eight stars in the sample, albeit intending them to be representative.  In a more comprehensive treatment, these would be randomized and multiple mission simulations—with and without the application of the methodology presented in this paper—would be run to gain statistics and make a quantitative estimation of its efficacy for the three major coronagraph types.

The research was carried out at the Jet Propulsion Laboratory, California Institute of Technology, under a contract with the National Aeronautics and Space Administration. © 2010 California Institute of Technology. Government sponsorship acknowledged.  We thank Joseph Catanzarite, Stephen Unwin, Stephen Edberg, David Meier, Stuart Shaklan, Wesley Traub, Michael Shao, and James Marr for interesting conversations and constructive comments.